\documentclass{emulateapj}

\slugcomment{Accepted for publication in ApJ}

\shorttitle{Jet Power in Radio Galaxies}
\shortauthors{Godfrey \& Shabala}

\begin{document}

\title{AGN Jet Kinetic Power and the Energy Budget of \\ Radio Galaxy Lobes}

\author{L. E. H. Godfrey\altaffilmark{1}, S. S. Shabala\altaffilmark{2}}

\email{L.Godfrey@curtin.edu.au}

\altaffiltext{1}{International Centre for Radio Astronomy Research, Curtin University, GPO Box U1987, Perth, WA, 6102, Australia}
\altaffiltext{2}{School of Mathematics and Physics, Private Bag 37, University of Tasmania, Hobart, TAS 7001, Australia}

\begin{abstract}

Recent results based on the analysis of radio galaxies and their hot X-ray emitting atmospheres suggest that non-radiating particles dominate the energy budget in the lobes of FRI radio galaxies, in some cases by a factor of more than 1000, while radiating particles dominate the energy budget in FRII radio galaxy lobes. This implies a significant difference in the radiative efficiency of the two morphological classes. To test this hypothesis, we have measured the kinetic energy flux for a sample of 3C FRII radio sources using a new method based on the observed parameters of the jet terminal hotspots, and compared the resulting $Q_{\rm jet} - L_{\rm radio}$ relation to that obtained for FRI radio galaxies based on X-ray cavity measurements. Contrary to expectations, we find approximate agreement between the $Q_{\rm jet} - L_{\rm radio}$ relations determined separately for FRI and FRII radio galaxies. This result is ostensibly difficult to reconcile with the emerging scenario in which the lobes of FRI and FRII radio galaxies have vastly different energy budgets. However, a combination of lower density environment, spectral ageing and strong shocks driven by powerful FRII radio galaxies may reduce the radiative efficiency of these objects relative to FRIs and couteract, to some extent, the higher radiative efficiency expected to arise due to the lower fraction of energy in non-radiating particles. An unexpected corollary is that extrapolating the $Q_{\rm jet} - L_{\rm radio}$ relation determined for low power FRI radio galaxies provides a reasonable approximation for high power sources, despite their apparently different lobe compositions.

\end{abstract}

\keywords{galaxies: active -- galaxies}

\section{Introduction}

Measuring the kinetic power of extragalactic jets has application in two important areas of astrophysics: (1) determining the Active Galactic Nucleus (AGN) kinetic luminosity function and its evolution --- an important factor in the study of radio galaxy feedback \citep{croton06, shabala09, fanidakis11}; and (2) assessing the contribution of black hole spin in the production of extragalactic jets \citep[e.\,g.][and references therein]{daly09}.  Furthermore, accurate estimates of jet kinetic energy flux can provide an important constraint in assessing X-ray emission models of kpc-scale quasar jets \citep{godfrey12}. Accurate measurement of jet power in radio galaxies is key to quantifying their effect on galaxy evolution. For example, \citet{rawlings04} and \citet{shabala11} showed that bow shocks driven by the high power radio sources can suppress star formation in not only the AGN host, but also in other nearby galaxies. More importantly for the current work, the ratio of radio luminosity ($L_{\rm radio}$) to jet power ($Q_{\rm jet}$), also known as the radiative efficiency, is sensitive to the division of energy between radiating and non-radiating particle populations, and therefore may be used to investigate the lobe energetics. However, measuring the kinetic power of radio galaxies has proven to be a very difficult problem. 
The lack of reliable empirical methods to measure the kinetic power of AGN jets has resulted in the widespread use of a model-dependent predictor of jet power derived by \citet{willott99}, based on synchrotron minimum energy calculations in combination with the self-similar model of radio galaxy evolution \citep{falle91, kaiser97}. \citet{willott99} obtain an expression for the jet power $Q_W$ (``W" for Willott) in terms of the 151 MHz radio luminosity 
\begin{equation} \label{eqn:willott}
Q_W \approx f^{3/2} ~ 3 \times 10^{38} \left(  \frac{L_{151}}{10^{28}~{\rm W~Hz^{-1}~sr^{-1}}} \right)^{6/7} ~ {\rm W}
\end{equation}
where $Q_W$ is time averaged kinetic power of a source with radio luminosity $L_{151} = F_{\rm 151} D_L^2$, and $f$ is a parameter accounting for systematic error in the model assumptions. These model assumptions include, among other things, the fraction of energy in non-radiating particles, the low frequency cutoff in the synchrotron spectrum, and departures from minimum energy. It is argued by \citet{willott99} that $1 \leq f \leq 20$, implying a systematic uncertainty of 2 orders of magnitude in jet power for a given radio luminosity, owing to the $f^{3/2}$ dependence. The Willott et al. jet power relation is widely used to estimate the mechanical output from AGN based on a single low frequency luminosity measurement, assuming that the value of $f$ is constant (typically of order 10 - 20) across the entire population of radio galaxies \citep[e.\,g.][]{hardcastle07, martinez-sansigre11, fernandes11, cattaneo09}.  The value of $f$ is often calibrated against FRI radio galaxies for which the jet power has been determined based on the observed X-ray cavities \citep[e.\,g.][]{rafferty06}. However, this procedure of calibrating the value of $f$ for FRII radio galaxies based on measurements of FRI radio galaxies may not be appropriate because of their vastly different energy budgets: non-radiating particles are thought to dominate the energy budget of FRI radio lobes by a factor of $\gg 100$ in some cases \citep{croston03, croston08, birzan08}, while in FRII radio galaxies, radiating particles are thought to dominate the energy budget \citep{croston04, croston05, belsole07}, indicating that significantly different values of $f$ should apply to the different morphological classes. Indeed, it appears that vastly different values of $f$ apply to FRI radio galaxies with different evolutionary histories \citep{cavagnolo10}. Moreover, the self-similar model of radio source evolution, on which Equation \ref{eqn:willott} is based, does not strictly apply to FRI radio galaxies. 

Not only is the normalisation of the \citet{willott99} relation highly uncertain, so too is the exponent. \citet{willott99} use the OII narrow line luminosity ($L_{\rm OII}$) as a proxy for jet power, and argue that Equation \ref{eqn:willott} is valid because the power law exponent matches that of the $L_{\rm OII} - L_{151}$ correlation that they find in their sample. However, the slope of the $L_{OII} - L_{151}$ correlation is highly uncertain, and strongly depends on the sample involved: \citet{hardcastle09} find $L_{OII} \propto L_{178}^{1.02 \pm 0.2}$, while \citet{fernandes11} find $L_{OII} \propto L_{178}^{0.52 \pm 0.1}$. Moreover, $L_{OII}$ is expected to exhibit a fairly weak dependence on accretion power, and the relationship is likely to be non-linear in general \citep{tadhunter98, hardcastle09}. \citet{hardcastle09} find that the correlation between $L_{OII}$ and accretion related X-ray emission is not significant after the common correlation with redshift is accounted for, however, \citet{shabala12} find a correlation between jet power and narrow line luminosity for a sample of flat-spectrum radio quasars. Finally, \citet{osullivan11} showed that when the minimum energy calculations are performed correctly, the exponent in Equation \ref{eqn:willott} depends on the spectral index of the source, which in turn is a function of source age (Shabala \& Godfrey 2012, hereafter Paper 2; Blundell et al. 1999). In the present work, we seek to address these outstanding questions regarding the $Q_{\rm jet} - L_{\rm radio}$ relation in FRII radio galaxies using an empirical means to determine the jet power.

Jet power in FRII radio galaxies and quasars can be estimated directly from measurements of the hotspot size and equipartition magnetic field strength, along with a number of reasonable assumptions regarding the hotspot plasma. In this paper we provide a derivation and analysis of this technique, and investigate the relationship between jet kinetic power and monochromatic radio luminosity for a sample of high power FRII radio galaxies. We compare the resulting $Q_{\rm jet} - L_{\rm radio}$ relation to the equivalent relation obtained for FRI radio galaxies based on X-ray cavity measurements, as well as that derived by Willott et al. (Equation \ref{eqn:willott}). 

In Section 2 we provide a derivation of the jet power measurement technique based on the parameters of the terminal hotspots. In Section 3 we describe the sample selection and analysis. In Section 4 we present our results, including a comparison of the $Q_{\rm jet} - L_{\rm radio}$ relations for FRI and FRII radio galaxies. In Section 5 we discuss the implications of our findings and in Section 6 we present our conclusions. 

Throughout this paper, the spectral index ($\alpha$) is defined such that $S_\nu \propto \nu^{-\alpha}$, and we adopt the following values for cosmological parameters: $H_0 = 71$~km~$s^{-1}$~Mpc$^{-1}$, $\Omega_m = 0.27$, $\Omega_\Lambda = 0.73$. Note that, following convention, we define the radio luminosity as as $L_{\nu} = S_\nu D_L^2$, where $S_\nu$ is the flux density, and $D_L$ is the luminosity distance.

\section{jet power in FRII radio galaxies based on measurements of the terminal Hotspots}  \label{sec:theoretical_background}

Consider a uniform jet of area $A$, particle energy density $\epsilon$, pressure $p$, mass density $\rho$,  relativistic enthalpy $w = \epsilon + p + \rho c^2$, magnetic field components perpendicular and parallel to the flow direction $B_{\perp}$ and $B_{||}$, speed $\beta c$ and corresponding bulk Lorentz factor $\Gamma$. The flux of energy ($F_E$) and momentum ($F_M$) along the jet are \citep[e.\,g.][]{double04}:
\begin{eqnarray}
F_E &=& A \Gamma^2 \beta c \left( w + \frac{B_{\perp}^2}{4 \pi}  \right) \qquad  \label{eqn:F_E} \\
F_M &=& A \left[ \Gamma^2 \beta^2 \left(  w + \frac{B_{\perp}^2}{4 \pi}   \right)  + p + \left( \frac{B_{\perp}^2 - B_{||}^2}{8 \pi} \right) \right] \qquad \label{eqn:F_M}
\end{eqnarray} 
In a highly relativistic jet with $\Gamma \gg 1$, the energy flux $F_{\rm E, jet}$ is simply related to the momentum flux $F_{\rm M, jet}$ via (see the Appendix) 
\begin{eqnarray}
F_{\rm E, jet} &\approx& c \times  F_{\rm M, jet}. 
\end{eqnarray}
Conservation of momentum between the jet and hotspot then implies
\begin{eqnarray}
F_{\rm E, jet} &\approx& c \times F_{\rm M, hs} \label{eqn:F_E_vs_F_M}
\end{eqnarray}
where $F_{\rm M, hs}$ is the momentum flux in the hotspot. This equality holds regardless of assumptions about the jet characteristics such as its composition or the ratio of magnetic to particle energy densities in the jet or hotspot. Therefore, if the general principle of conservation of momentum applies between the jet and hotspot, we can estimate the jet kinetic luminosity simply by calculating the momentum flux in the hotspot. To do so, we assume that the lepton population in the hotspot is ultra-relativistic ($\epsilon_{e^\pm} = 3 p_{e_\pm}$) and that the proton population in the hotspot, if it exists, is at best mildly relativistic, and can be approximated as a thermal gas ($\epsilon_p = (3/2) p_p$) so that the hotspot enthalpy density, $w$, is parametrized as follows
\begin{equation}
w = \epsilon_{e^\pm} \left( \frac{4}{3} + \frac{5}{3} \frac{\epsilon_p}{\epsilon_{e^\pm}} + \frac{\rho c^2}{\epsilon_{e^\pm} } \right) \label{eqn:w}
\end{equation}
We further assume that in the hotspot, the magnetic field is aligned perpendicular to the jet direction, i.\,e. $B_\perp = B$ and $B_{||} = 0$, consistent with radio polarisation maps of hotspots. The hotspot momentum flux is then:
\begin{eqnarray}
F_{\rm M, hs} &=& A~ c~ \epsilon_{e^\pm} \left[ \Gamma^2 \beta^2 \left( \frac{4}{3} + \frac{5}{3} \frac{\epsilon_p}{\epsilon_{e^\pm}} + \frac{\rho c^2}{\epsilon_{e^\pm} } \right)  + \left( \frac{1}{3} + \frac{2}{3} \frac{\epsilon_p}{\epsilon_{e\pm}} \right) \right]  \nonumber \\
&+& A~ c~ \epsilon_{B} \left[ 1 + 2 \Gamma^2 \beta^2 \right]  \nonumber
\end{eqnarray}
Let $B_{\rm eq}$ be the equipartition magnetic field strength, calculated using standard expressions \citep[e.\,g.][]{worrall09}, assuming negligible energy density in non-radiating particles. Without loss of generality, we can write
\begin{equation}
\epsilon_{e^\pm} = \frac{B^2_{\rm eq}}{8 \pi} \left( \frac{B}{B_{\rm eq}} \right)^{-(1+\alpha)}
\end{equation}
and then express the jet energy flux in terms of $B_{\rm eq}$,
\begin{equation} \label{eqn:Q_hs_b_eq_0}
Q_{\rm HS} = A~c~\frac{B^2_{eq}}{8 \pi} \times g  
\end{equation}
where
\begin{eqnarray} \label{eqn:g-factor}
&g& \left( \alpha, \beta, \frac{\epsilon_p}{\epsilon_{e^\pm}}, \frac{\rho c^2}{\epsilon_{e^\pm}}, \frac{B}{B_{\rm eq}} \right) =     \left(  1 + 2 \Gamma^2 \beta^2 \right)  \left( \frac{B}{B_{\rm eq}} \right)^2 \nonumber \\ &+& \left[ \Gamma^2 \beta^2 \left( \frac{4}{3} + \frac{5}{3} \frac{\epsilon_p}{\epsilon_{e^\pm}} + \frac{\rho c^2}{\epsilon_{e^\pm} } \right)  + \left( \frac{1}{3} + \frac{2}{3} \frac{\epsilon_p}{\epsilon_{e\pm}} \right) \right] \left( \frac{B}{B_{\rm eq}}  \right)^{-(1+\alpha)} \nonumber \\
\end{eqnarray}

\subsection{Empirical determination of $g$, the normalization factor} \label{sec:empirical_g}

In this section, we empirically determine the value of $g$ in equation \ref{eqn:Q_hs_b_eq_0} by applying the hotspot method to sources with independent jet power measurements. 

The prototypical FRII radio galaxy Cygnus A is the prime candidate for calibrating the normalisation factor, since its hotspot parameters are well determined and the jet power has been independently measured using a variety of methods. \citet{wilson06} estimated the jet power of Cygnus A to be $Q \gtrsim 1.2 \times 10^{46}$~erg~s$^{-1}$ based on analysis of the cocoon dynamics determined using \emph{Chandra} X-ray imaging spectroscopy. \citet{ito08} obtain a similar estimate of jet power ($Q = 0.4 - 2.6 \times 10^{46}$~ergs~s$^{-1}$), based on dynamical modelling of the source. An independent estimate of jet power in Cygnus A comes from \citet{lobanov98} who shows that frequency dependent shifts of the radio core enable a determination of the jet power \citep[see also][]{shabala12}, and when applied to the case of Cygnus A, gives $Q \approx 6 \times 10^{45}$~erg~s$^{-1}$. \citet{rafferty06} estimate the jet power of Cygnus A from the X-ray cavity, and find $Q \approx 1.3 \times 10^{45}$~erg~s$^{-1}$, but more recent analysis incorporating the energy in shocks suggests $Q \gtrsim 5 \times 10^{45}$~erg~s$^{-1}$ (P. Nulsen, Priv. Comm. 2012). We estimate the jet power of Cygnus A using Equation~\ref{eqn:Q_hs_b_eq_0} along with the parameters for the terminal hotspots (hotspots A and D) given in \citet{wright04}. The derived jet power is $Q = g \times 4 \times 10^{45}$~ergs~s$^{-1}$, which implies $g \approx 1 - 2$ in this source. 

We can apply a similar analysis to the FRII radio galaxy 3C 401, as it has an independent jet power measurement from analysis of the associated X-ray cavity: \citet{rafferty06} estimate the jet power of 3C401 to be $Q \approx 7 \times 10^{44}$~erg~s$^{-1}$. This source does not appear in our sample because the beam size of the highest resolution map available to us is larger than $1\%$ of the source linear extent, and therefore, the hotspots may not be adequately resolved (see Section \ref{sec:sample}). The hotspots in this source are faint relative to the surrounding lobe emission, making the determination of hotspot parameters difficult. Despite these problems, based on the hotspot sizes and flux density given by \citet{mullin08}, we obtain a hotspot jet power estimate of $Q = g \times 3.2 \times 10^{44}$~ergs~s$^{-1}$, which implies $g \sim 2$. 

Finally, we note that \citet{daly12} estimate the jet power for a sample of 31 high power FRII radio galaxies, using the expression $Q = 4 p V / \tau$, where $p$ is the lobe pressure calculated using minimum energy arguments, $V$ is the lobe volume assuming cylindrical symmetry, and $\tau$ is the spectral age of the source. The \citet{daly12} sample includes four sources from our sample (3C 55, 3C 244.1, 3C 289 and 3C 337). For these four sources, the average value of $g$ required to match the jet power values of \citet{daly12} is $g = 2 \pm 1$. 

The low value derived for the normalisation factor ($g \sim 2 $) indicates that the hotspot plasma is close to equipartition conditions, consistent with the results of inverse Compton modelling in hotspots \citep{hardcastle04} and lobes \citep{croston05} of FRII radio galaxies. 

\subsection{Expectations for $g$}  \label{sec:g_theoretical}

Here we argue that the value for the normalisation factor derived from observations in the previous section is consistent with expectations, by combining constraints on the various parameters involved in Equation \ref{eqn:g-factor}. 

We first consider the ratio $\frac{\rho c^2}{\epsilon_{e^\pm}}$. We can constrain this ratio, assuming at most one proton per radiating lepton, via an estimate of the mean electron Lorentz factor $\langle \gamma \rangle$ in the hotspot, since
\begin{eqnarray} \label{eqn:rhoc2}
\frac{\rho c^2}{\epsilon_{e^\pm}} &=& \frac{\left( 1 + \frac{n_p m_p}{n_{e^\pm} m_{e}}  \right)}{\langle \gamma  \rangle - 1} \lesssim \frac{1837}{\langle \gamma  \rangle} 
\end{eqnarray}
For a power-law electron energy distribution of the form $N(\gamma) = k_e \gamma^{-a}$, the mean electron Lorentz factor $\langle \gamma  \rangle \approx \gamma_{\rm min} \left( \frac{a-1}{a-2} \right)$ for $a>2$ and $\langle \gamma  \rangle = \gamma_{\rm min} \ln \left( \frac{\gamma_{\rm max}}{\gamma_{\rm min}} \right)$ for $a=2$.  In each case where flattening of the hotspot radio spectrum has been directly observed, estimates of $\gamma_{\rm min}$ are in the order of several hundred: PKS~1421--490, $\gamma_{\rm{min}} \sim$ 650 \citep{godfrey09}; Cygnus A,
$\gamma_{\rm{min}} \sim$ 300 - 400 \citep{carilli91, lazio06, hardcastle01b};
3C295, $\gamma_{\rm{min}} \sim$ 800 \citep{harris00, hardcastle01b}; 3C123,
$\gamma_{\rm{min}} \sim$ 1000 \citep{hardcastle01a, hardcastle01b}.  In addition, \citet{hardcastle01b} inferred a cutoff Lorentz factor $\gamma_{\rm min} \sim 500$ in 3C 196 by synchrotron self-Compton modelling of the hotspot spectral energy distribution. All of the above listed $\gamma_{\rm min}$ estimates appear to be distributed around a value of order $\gamma_{\rm min} \sim 500$, and this value of the minimum Lorentz factor may arise naturally through the dissipation of bulk kinetic energy in the hotspots \citep{godfrey09}. We therefore assume $\gamma_{\rm min} \sim 500$, and with typical electron energy index $a \sim 2.2 - 2.6$, Equation \ref{eqn:rhoc2} implies $\rho c^2/\epsilon_{e^\pm} \lesssim 1$. 

We now consider the ratio $\epsilon_p/\epsilon_{e^\pm}$. \citet{hardcastle04} argue, based on synchrotron self Compton modelling of hotspot X-ray emission, that an energetically dominant proton population is disfavoured, such that $\epsilon_p/\epsilon_{e^\pm} \lesssim 1$.
The lobes of FRII radio galaxies are inflated by backflow of hotspot plasma. It has been shown that $\epsilon_p/\epsilon_{e^\pm} \lesssim 1$ in the lobes of FRII radio galaxies \citep{croston04, croston05, belsole07}, further suggesting that the ratio $\epsilon_p/\epsilon_{e^\pm} \lesssim 1$ in hotspots.

\begin{figure}[!h] 
\epsscale{1.0}
\plotone{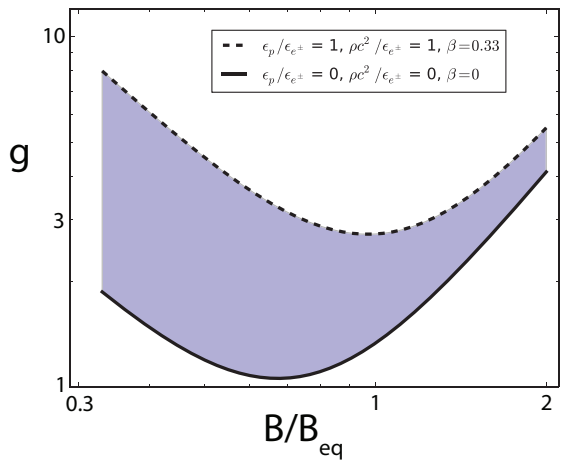}
\caption{The normalization factor $g$ (see Equation \ref{eqn:Q_hs_b_eq_0}) as a function of $B/B_{\rm eq}$ for a range of assumed states of the hotspot plasma. Without loss of generality, we have defined $B_{\rm eq}$ to be the equipartition magnetic field strength calculated assuming negligible energy density in non-radiating particles. The solid curve corresponds to a case in which the jet is purely leptonic and the hotspot plasma has a non-relativistic velocity. These assumptions correspond to the minimum possible jet power for a given hotspot size and luminosity, with $g = 1.06$, at $B/B_{\rm eq} \approx 2/3$. The dashed curve corresponds to a hotspot in which the radiating and non-radiating particle populations have equal energy density, the rest mass energy density is equal to that of the radiating particle energy density, and the hotspot maintains a mildly relativistic velocity of 0.33 c. \label{fig:g_function}}
\end{figure}

We next consider the ratio $\frac{B}{B_{\rm eq}}$. \citet{hardcastle04} argue, on the basis of synchrotron self Compton modelling of X-ray emission in a large sample of FRII radio galaxy hotspots, that magnetic field strengths are close to the equipartition estimates; that is $\frac{B}{B_{\rm eq}} \approx 1$. 

Finally, we consider the bulk velocity of the hotspot plasma, $\beta$. The post-shock velocity following a normal shock in an unmagnetised plasma with relativistic equation of state is $\frac{1}{3}c$.  \citet{dennett-thorpe97} discovered a hotspot spectral index asymmetry in the sense that hotspots fed by the approaching jet have a flatter spectral index than hotspots fed by the receding jet. This observation can be explained if the hotspots are moderately Doppler beamed and have curved spectra, with higher frequencies corresponding to a steeper spectrum. The observed hotspot spectral index asymmetry requires only moderate velocity in the hotspot regions, with $\beta \gtrsim 0.3$ \citep{dennett-thorpe97}. 

Given the constraints on hotspot parameters discussed above, we plot the function $g$ to determine physically realistic values (Figure \ref{fig:g_function}). 
In this Figure we plot $g$ as a function of $B/B_{\rm eq}$ for $1/3 < \frac{B}{B_{\rm eq}} < 2$, $0 < \frac{\rho c^2}{\epsilon_{e^\pm}} < 1$, $\epsilon_p/\epsilon_{e^\pm} \lesssim 10$, $0 < \beta < 0.33$, and $\alpha = 0.5$. It is clear that the emprically derived value of $g \sim 1 - 2$ is consistent with expectations. We note that $g \approx 3$ for a hotspot near minimum energy with a proton/electron composition, $\beta \sim 1/3$ and $\epsilon_e \approx \epsilon_p \approx \rho c^2$, while $g \approx 1.5$ for a purely leptonic hotspot near equipartition conditions and $\beta \sim 1/3$.

\section{Sample Selection and Data Analysis} \label{sec:sample}

We select a subset of the complete, flux limited sample of 3CRR FRII radio sources with redshifts $z<1$ \citep{mullin08}. In order to limit the effects of observing resolution, we excluded objects for which there are fewer than 100 beam widths across the source in the highest resolution maps available to us. Hotspot sizes are typically 1 percent of the source size \citep{hardcastle98}, and therefore a lower limit of 100 beams along the source ensures that the beam size is smaller than, or approximately equal to the hotspot size. A total of 62 sources met this criteria. We further restricted our sample to include only those sources for which a clear jet termination could be determined for at least one side of the source, resulting in a sample size of 30. Using the radio maps available from the 3CRR FRII online database, we independently measured the hotspot sizes and flux densities in a consistent manner: we fit a 2D-Gaussian function to each hotspot using the CASA task IMFIT, but only included in the fit those pixels with a value greater than half the hotspot peak value. This typically resulted in $50 - 80$ pixels being used in the fit. We do not attempt to model or subtract the background contributed by the surrounding lobe emission, but we have mitigated the effect of the lobe emission by restricting the model-fitting routine to include only the bright part of the hotspot. Also, our sample was chosen to include only those hotspots that were well resolved from the surrounding lobe emission, and we expect that for the majority of hotspots in our sample, the lobe emission has a negligible impact on our derived hotspot size/flux density. For each source, we used the highest resolution FITS image available from the online database. The hotspot jet powers derived in this way are on average $50\%$ greater than the values determined using the measured hotspot size and flux density given by \citet{mullin08}.  In at least one instance, the fitting routine gave unreasonable results, and we then estimated the hotspot region from the map, and extracted the flux using an elliptical aperture of the same dimensions as the hotspot dimensions estimated from cross-sectional profiles of the hotspot.

As with \citet{mullin08}, we have not quoted errors for hotspot flux density or size since the dominant source of error comes from the ambiguity in defining the hotspot region, and is therefore, to some extent, subjective. However, we have minimised the ambiguity by restricting our sample to include only those sources for which the beam width is less than, or comparable to the expected hotspot size \citep{mullin08}, and therefore expect our derived hotspot powers to be reliable estimates, free of large systematic bias.

\begin{figure}[!ht]
\epsscale{1.0}
\plotone{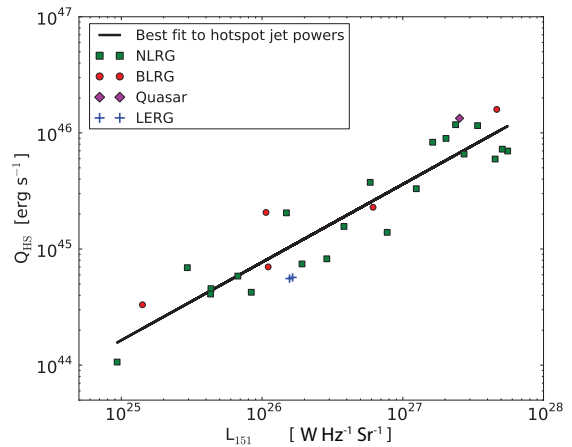}
\caption{Jet power, determined using the hotspot method, as a function of 151 MHz monochromatic radio luminosity. 
The solid black line is the best fit relation to our data, which are plotted without multiplication by the normalization factor (effectively assuming $g=1$). \label{fig:Q_hs}}
\end{figure}

\newpage

\section{Results}

For each hotspot, we compute the equipartition magnetic field strength $B_{\rm eq}$ using standard expressions \citep{worrall09} assuming negligible energy density in non-radiating particles. We assume a power-law electron energy distribution of the form $N(\gamma) = k_e \gamma^{-a}$ with $a = 2.2$ between a minimum Lorentz factor $\gamma_{\rm min} = 500$ \citep[see][]{godfrey09} and maximum Lorentz factor $\gamma_{\rm max} = 10^5$. For each hotspot, we then calculate the jet power using Equation \ref{eqn:Q_hs_b_eq_0} with $g=2$. We sum the power derived from the hotspots on each side of the source to obtain the total source power. In Section \ref{sec:calorimeters}, we discuss the reliability of jet power measurements from individual hotspots, and find that typically the jet power estimated for a pair of hotspots in the same source agree to within a factor of 2. In a few cases, only one hotspot could be used for a reliable jet power estimate, and in that case, we multiplied the derived value by a factor of 2, to account for the power in the oppositely directed jet. The sample and derived jet powers are given in Table \ref{table:sample}. We perform least-squares minimisation in log space to fit a power law of the form $Q = A L_{151}^B$ to the data, and find a best-fit relation
\begin{equation} \label{eqn:our_best_fit}
Q_{\rm FRII} = g \times (1.5 \pm 0.5)  \times 10^{44} \left(  \frac{L_{151}}{10^{25}~{\rm W}~{\rm Hz}^{-1}~{\rm sr}^{-1}} \right)^{0.67 \pm 0.05}~.
\end{equation}
Calculating the hotspot jet powers using the measured hotspot parameters tabulated by \citet{mullin08} results in jet powers that are on average a factor of 1.5 lower. Assuming $\gamma_{\rm min} = 10$ rather than $\gamma_{\rm min} = 500$ increases the normalisation of the best-fit curves by a factor of approximately 1.5. The assumed hotspot spectral index affects the slope of the derived best-fit relation, however the effect is comparable to the statistical uncertainty in the slope due to the scatter in the correlation. For example, assuming that the hotspot spectral index is $\alpha = 0.8$ instead of $\alpha = 0.6$ results in a marginally flatter best-fit relation, with an exponent of $0.62 \pm 0.05$.

\subsection{Comparison with the $Q_{\rm jet} - L_{151}$ relation for FRI radio galaxies}

A number of authors have investigated the relationship between AGN jet power and radio power in nearby low-luminosity sources using X-ray cavity measurements; the most widely discussed recent work being that of \citet{cavagnolo10} \citep[see also][]{birzan08, osullivan11}. Studies of jet kinetic power based on X-ray cavity measurements are inherently limited to relatively nearby, low power objects, typically of FRI morphology. We seek to compare our results to the $Q_{\rm jet} - L_{\rm 151}$ relation obtained for FRI radio galaxies. To do so, we have used the X-ray cavity jet power estimates compiled by \citet{cavagnolo10} along with 151 MHz radio luminosities extrapolated from low frequency measurements, assuming a spectral index $\alpha = 0.8$. The low-frequency luminosities for the Cavagnolo sample lie between $200-400$ MHz, and are mostly at 327MHz. This is sufficiently close to 151 MHz for the assumed value of spectral index to be largely insignificant: a departure in the assumed spectral index of $\Delta \alpha = 0.3$ results in only a $\sim 20 \%$ error in $L_{\rm 151}$. We have converted the best-fit relation of \citet{cavagnolo10} to one involving $L_{\rm 151}$:
\begin{equation} \label{eqn:Q_FRI}
Q_{\rm FRI} = 5^{+2}_{-1} \times 10^{44} \left( \frac{L_{151}}{10^{25}~{\rm W}~{\rm Hz}^{-1}~{\rm sr}^{-1}} \right)^{0.64 \pm 0.09}  
\end{equation}

The exponent in the above relation for FRI radio galaxies is in excellent agreement with the exponent determined for our sample of high power FRII sources (Equation \ref{eqn:our_best_fit}). We plot Equation \ref{eqn:Q_FRI} in Figure \ref{fig:Q_hs} along with the data for the \citet{cavagnolo10} sample of low-luminosity radio galaxies (red points), and the data for our sample of FRII radio galaxies (green squares) using the assumption $g=2$ (see Section \ref{sec:empirical_g}). We find no evidence for a significant offset between the FRI and FRII $Q_{\rm jet} - L_{\rm radio}$ relations: the normalisations formally agree if $g \gtrsim 2$, which is entirely consistent with our analysis in Sections \ref{sec:empirical_g} and \ref{sec:g_theoretical}. \citet{daly12} estimate the jet power for a sample of 31 high luminosity FRII radio galaxies using the expression $Q = 4 p V / \tau$, where $p$ is the lobe pressure calculated using minimum energy arguments, $V$ is the lobe volume assuming cylindrical symmetry, and $\tau$ is the spectral age of the source. In agreement with the results presented here, Daly et al. also found that the $Q_{\rm jet} - L_{\rm radio}$ relation for luminous FRII radio galaxies is in broad agreement with an extrapolation of the one given by \citet{cavagnolo10} for FRI radio galaxies.

\begin{figure}[!ht]
\epsscale{1.0}
\plotone{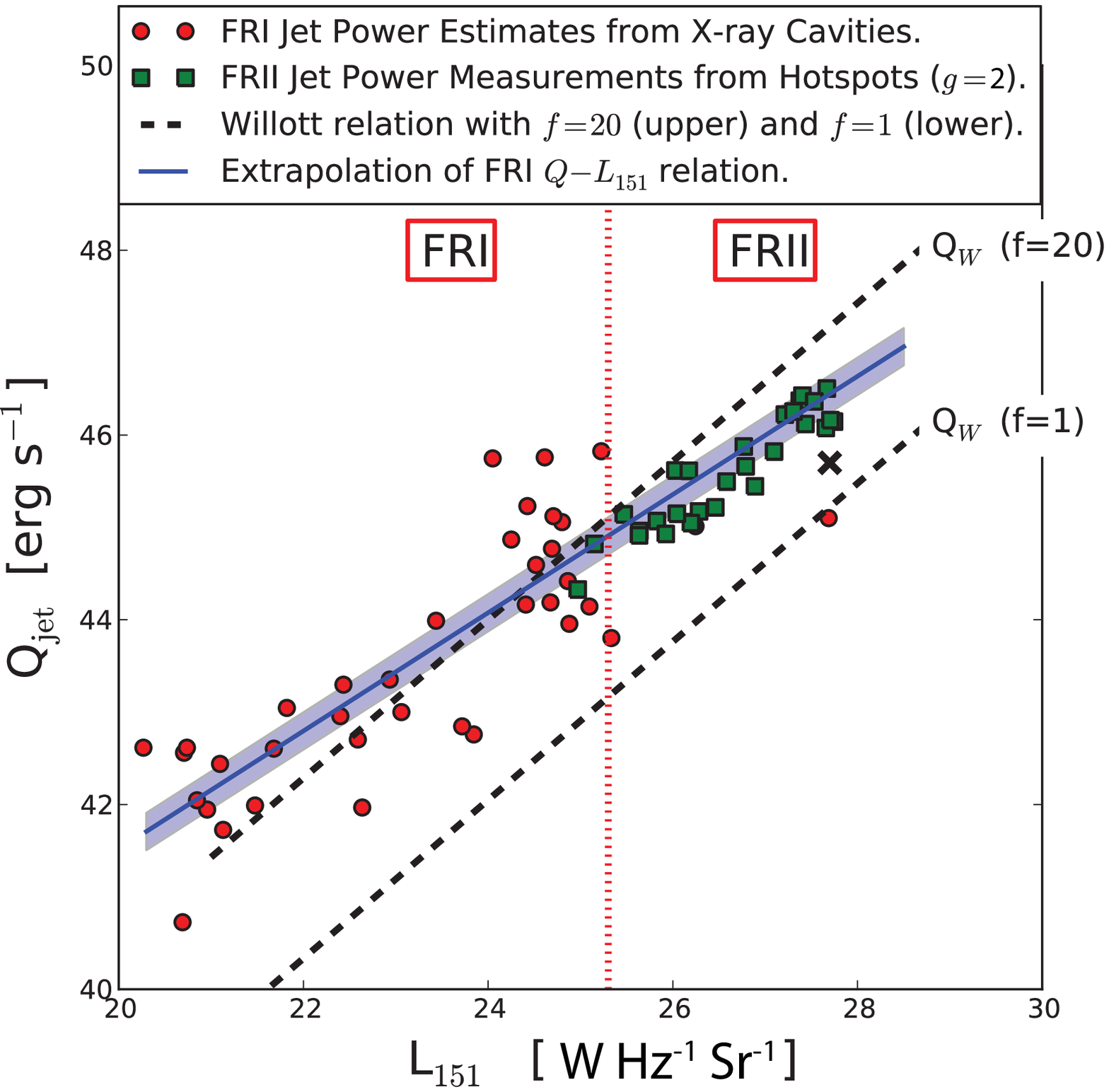}
\caption{Comparison of the $Q_{\rm jet} - L_{151}$ relations for FRI and FRII radio galaxies. Here we plot the data and best fit relation from \citet{cavagnolo10} (red points and blue solid line). The shaded area illustrates uncertainty in the normalisation of the FRI best fit relation. We also plot the model of \citet{willott99} with $f=20$ (upper-most black dashed line) and $f=1$ (lower-most black dashed line). We plot the FRII jet power measurements (green squares) which have been derived using the hotspot method assuming $g=2$ (see Sections \ref{sec:empirical_g} and \ref{sec:g_theoretical}). Note that the minimum allowed value is $g=1.06$. The black cross marks the location of Cygnus A and is clearly an outlier when compared to our sample of FRII radio galaxies. This is due to the high density environment into which Cygnus A expands, resulting in ``environmental boosting" of its radio luminosity \citep[][see also Section \ref{sec:predicted_offset}]{barthel96}. \label{fig:Q_hs_with_cavagnolo}}
\end{figure}

Our result, the broad agreement between the FRI and FRII $Q_{\rm jet} - L_{\rm radio}$ relations, appears at odds with the emerging scenario in which the fraction of energy in non-radiating particles differs greatly between these two classes of radio galaxy. However, as we discuss in Section \ref{sec:predicted_offset}, differences in the age and environment for the two samples used in this study, as well as a possible difference in the fraction of energy associated with shocks, will counteract the offset between the $Q_{\rm jet} - L_{\rm 151}$ relations expected to arise due to the differing energy budgets of the radio lobes.

\begin{figure}[!ht]
\epsscale{1.0}
\plotone{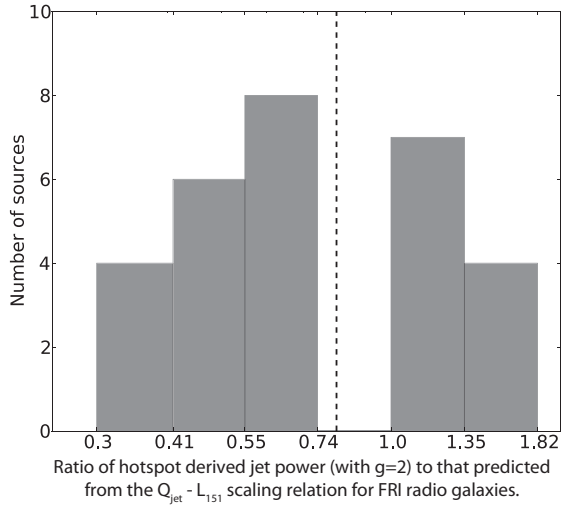}
\caption{Histogram of the ratio between hotspot jet power (with $g=2$) and the jet power calculated from Equation \ref{eqn:Q_FRI}, the $Q_{\rm jet} - L_{\rm 151}$ scaling relation for FRI radio galaxies. The mean of this distribution (0.8) is illustrated by the dashed line. It is clear that given $g \approx 2$ as derived in section \ref{sec:empirical_g}, there is no evidence for a substantial offset between the FRI and FRII $Q_{\rm jet} - L_{\rm radio}$ relations.  \label{fig:FRI_comparison_histogram}}
\end{figure}

\section{Discussion}

\subsection{Hotspots as calorimeters} \label{sec:calorimeters}

\begin{figure}[!ht]
\plotone{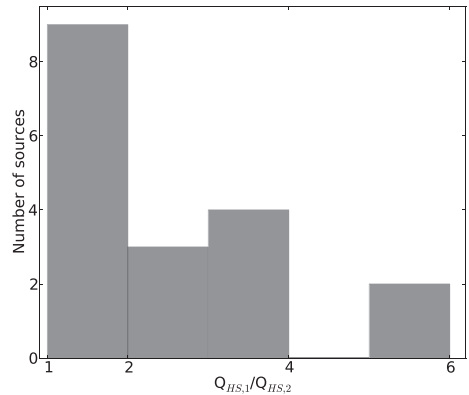}
\caption{Histogram of the ratio of jet power derived from the two hotspots at either end of the source. The median of the distribution is 2.0 and standard deviation is 1.4.  \label{fig:histogram}}
\end{figure}

Hotspots of FRII radio galaxies are thought to be variable on short timescales \citep{laing89, saxton02, saxton10}, and as such, caution must be exercised when interpreting the derived jet power for individual objects. However, provided that the general principle of conservation of momentum applies between jet and hotspot, on a population basis we expect this method to be a reliable estimator of jet power, and in particular, may be used to investigate the $Q_{\rm jet} - L_{\rm radio}$ relation at high radio luminosities. More than half of the sources in our sample have two hotspots, one at each end of the source, that enable jet power estimates. We can test the reliability of the hotspot jet power method by calculating, for each source, the ratio of jet power determined for the two hotspots. Figure \ref{fig:histogram} is a histogram showing the distribution of this $Q_{\rm hs}$ ratio. More than half the sample have $Q_{\rm jet}$ estimates from both hotspots that agree to within a factor of two. The largest discrepancy between hotspot measurements is approximately a factor of 5.

\subsection{Predicted offset between the $Q_{\rm jet} - L_{\rm radio}$ relations for FRI and FRII radio galaxies}  \label{sec:predicted_offset}

\citet{osullivan11} revised the analysis of \citet{willott99} to account for a different minimum energy formalism. In particular, these authors pointed out that a large fraction of the synchrotron-emitting electron population may radiate below the frequency cutoffs assumed by \citet{willott99}. Because of this, it makes more sense to recast the minimum energy argument in terms of cutoff Lorentz factors of the electron energy distribution. An additional advantage of this approach is that it allows the $Q_{\rm jet} - L_{\rm radio}$ relation to be expressed in terms of the lobe spectral index. Below, we briefly recount the \citet{osullivan11} analysis.

Let $u_{\rm min}$ be the minimum energy of a synchrotron emitting source of a given volume and luminosity, $k$ the ratio of energy in non-radiating particles to the energy in radiating particles, and $B_{\rm me}$ the minimum energy magnetic field strength. In the model of \citet[][]{willott99} the jet power is shown to be $Q_{\rm jet} \propto u_{\rm min}^{3/2}$ (see their Equation 11). The minimum energy $u_{\rm min}$ is related to the minimum energy magnetic field stregnth via $u_{\rm min} \propto B_{\rm me}^2$. \citet{worrall06} provide an expression for the minimum energy magnetic field strength in terms of the observed source luminosity and volume, as well as the high and low energy cutoff in the electron energy distribution, which has the form $B_{\rm me} \propto \left( 1 + k  \right)^{\frac{1}{3 + \alpha}} L_{151}^{1/(3+\alpha)}$. Combining these expressions
\begin{equation} \label{eqn:k_dependence}
Q_{\rm jet} \propto \left( 1 + k  \right)^{\frac{3}{3 + \alpha}} L_{151}^{3/(3+\alpha)}
\end{equation}

\begin{deluxetable*}{lcccccccccc}
\tabletypesize{\scriptsize}
\tablecaption{Hotspot Properties and Derived Jet Power \label{table:sample}}
\tablewidth{0pt}
\tablehead{
Source & L$_{\rm 151}$ & \multicolumn{2}{c}{Hotspot B$_{\rm eq}$} & & \multicolumn{2}{c}{D$_{\rm hs}$\tablenotemark{a}} & & \multicolumn{2}{c}{Q$_{\rm jet}$\tablenotemark{b} (g=2)} & Q$_{\rm total}$\tablenotemark{c} (g=2) \\
\cline{3-4}  \cline{6-7} \cline{9-10} \\
 & $\times 10^{26}$ & North & South & & North & South & & North & South &  \\
 & W~Hz$^{-1}$~Sr$^{-1}$ & $\mu$G & $\mu$G & & kpc & kpc & & $\times10^{45}$ erg~s$^{-1}$ & $\times10^{45}$ erg~s$^{-1}$ & $\times10^{45}$ erg~s$^{-1}$ \\
}
\startdata
3C22	&	46.5	&	680	&	320	&	&	0.93	&	3.7	&	&	7.2	&	24	&	31	\\
3C33.1	&	1.07	&	24	&	10	&	&	19	&	19	&	&	3.5	&	0.65	&	4.1	\\
3C46	&	7.75	&	99	&	27	&	&	2.3	&	12	&	&	0.94	&	1.9	&	2.8	\\
3C55	&	55.6	&	290	&	160	&	&	1.7	&	4.4	&	&	4.5	&	9.5	&	14	\\
3C98	&	0.093	&	25	&	30	&	&	3.3	&	2.3	&	&	0.12	&	0.09	&	0.21	\\
3C109	&	6.15	&	93	&	191	&	&	2.5	&	2.3	&	&	0.96	&	3.6	&	4.6	\\
3C132	&	1.65	&	320	&	105	&	&	0.59	&	1.6	&	&	0.65	&	0.49	&	1.1	\\
3C184.1	&	0.43	&	33	&	68	&	&	5.0	&	2.2	&	&	0.49	&	0.43	&	0.91	\\
3C228	&	27.2	&	280	&	260	&	&	1.9	&	2.6	&	&	4.9	&	8.2	&	13	\\
3C234	&	2.88	&	170	&	120	&	&	0.81	&	2.2	&	&	0.36	&	1.3	&	1.7	\\
3C244.1	&	12.5	&	280	&	145	&	&	1.8	&	2.2	&	&	4.7	&	1.9	&	6.6	\\
3C284	&	1.92	&	30	&	81	&	&	6.0	&	2.7	&	&	0.60	&	0.89	&	1.5	\\
3C337	&	16.3	&	190	&	275	&	&	2.3	&	3.1	&	&	3.5	&	13	&	17	\\
3C340	&	23.7	&	91	&	250	&	&	9.3	&	3.1	&	&	12.9	&	10	&	24	\\
3C382	&	0.14	&	41	&	25	&	&	3.8	&	4.5	&	&	0.43	&	0.23	&	0.66	\\
3C33	&	0.43	&	-	&	160	&	&	-	&	0.95	&	&	-	&	0.41	&	0.82	\\
3C42	&	5.87	&	-	&	100	&	&	-	&	4.6	&	&	-	&	3.8	&	7.5	\\
3C223	&	0.67	&	-	&	17	&	&	-	&	10.5	&	&	-	&	0.58	&	1.2	\\
3C226	&	45.3	&	-	&	470	&	&	-	&	1.2	&	&	-	&	5.9	&	12	\\
3C263	&	25.3	&	-	&	500	&	&	-	&	1.7	&	&	-	&	13	&	27	\\
3C277.2	&	34.0	&	-	&	360	&	&	-	&	2.2	&	&	-	&	12	&	23	\\
3C289	&	50.9	&	-	&	950	&	&	-	&	0.67	&	&	-	&	7.2	&	14	\\
3C292	&	20.3	&	-	&	130	&	&	-	&	5.4	&	&	-	&	9.0	&	18	\\
3C300	&	3.82	&	-	&	93	&	&	-	&	3.2	&	&	-	&	1.6	&	3.1	\\
3C319	&	1.57	&	42	&	-	&	&	4.2	&	-	&	&	0.56	&	-	&	1.1	\\
3C349	&	1.49	&	-	&	130	&	&	-	&	2.7	&	&	-	&	2.1	&	4.1	\\
3C381	&	1.11	&	170	&	-	&	&	1.15	&	-	&	&	0.70	&	-	&	1.4	\\
3C452	&	0.84	&	27	&	69	&	&	8.0	&	1.7	&	&	0.87	&	0.26	&	1.1	\\
3C321	&	0.29	&	34	&	84	&	&	3.2	&	3.1	&	&	0.21	&	1.2	&	1.4	\\
\enddata
\tablenotetext{a}{Hotspot diameter, normal to jet direction.}
\tablenotetext{b}{Jet power derived from hotspot parameters using equation \ref{eqn:Q_hs_b_eq_0}, assuming the normalisation factor $g=2$, as derived in Section \ref{sec:empirical_g}.}
\tablenotetext{c}{Total source power (sum of north and south jet power, or twice the measured jet power if only one hotspot could be used), again, assuming $g=2$.}
\end{deluxetable*}

In Paper~2, we extend this analysis to show that the $Q_{\rm jet} - L_{\rm radio}$ relation is sensitive to a variety of radio source parameters, including lobe size and age. It is clear from Equation~\ref{eqn:k_dependence} however, that the $Q_{\rm jet} - L_{\rm radio}$ relation will be altered if either the the fraction of non-radiating particles $k$, or the spectral index $\alpha$ change (as expected from straightforward synchrotron ageing arguments, e.g. Alexander \& Leahy 1987).

We can estimate the expected offset between the $Q_{\rm jet} - L_{\rm radio}$ relations for FRI and FRII radio galaxies due to the presence of non-radiating particles using Equation \ref{eqn:k_dependence}. 
\citet{birzan08} estimate $k$ for a sample of mostly FRI radio galaxies by equating the internal lobe pressure with the external pressure, and requiring that the magnetic field strength be in energy equipartition with the particles. They find $k$ lies in the range $\sim 1 - 4000$, with a median value of $k \approx 180$. Similarly, \citet{croston08} find that the radiating material in a sample of FRI radio lobes is significantly under-pressured relative to the external environment, and infer the presence of non-radiating particle population that strongly dominates the lobe energy budget. In contrast, \citet{belsole07} find that the equipartition lobe pressures of FRII radio galaxies are typically close to pressure equilibrium with the external medium. Furthermore, \citet{croston05} find that the magnetic field strength in the lobes of FRII radio galaxies, determined via synchrotron and inverse Compton modeling of the radio to X-ray spectra, is close to the equipartition value determined from the radio data alone assuming negligible energy in non-radiating particles. They conclude that FRII lobes are unlikely to contain an energetically dominant population of non-radiating particles, because that would require the magnetic field energy to be matched to the energy of just the relativistic electron population rather than the energy of the entire particle population. The inferred difference in lobe energy budget is not altogether surprising, since FRII lobes are surrounded by bow shocks, and the jets are surrounded by cocoon plasma, making significant entrainment of ambient material rather more difficult than in FRI lobes, for which this is not the case.

If $k \sim 200$ in FRI radio galaxy lobes \citep[the median value from][]{birzan08} and $k \sim 1$ in FRII radio galaxy lobes, then naively applying Equation~\ref{eqn:k_dependence} to both morphological classes with $\alpha = 0.8$, we would expect the normalisation of the $Q_{\rm jet} - L_{\rm radio}$ relation for FRI radio galaxies to be greater than that of FRIIs by a factor of $\gtrsim 50$. This is clearly not observed (see Figures \ref{fig:Q_hs_with_cavagnolo} and \ref{fig:FRI_comparison_histogram}), and the close agreement between the $Q_{\rm jet} - L_{\rm radio}$ relations is puzzling, given such a large predicted offset.

However, the radio luminosity is affected by a number of other source parameters, including the density of the medium into which the radio source expands \citep{barthel96}. For a given radio luminosity, $Q_{\rm jet} \propto \rho^{-1/2}$ (Paper~2; Willott et al. 1999), where $\rho$ is a characteristic ambient gas density. FRI radio sources typically inhabit more dense environments than FRII radio sources \citep[e.\,g.][]{zirbel97, miller99}, so the environment dependence will reduce, to some extent, the predicted offset in radiative efficiency described above. 

We test this hypothesis by comparing the position of Cygnus A in the $Q_{\rm jet} - L_{\rm 151}$ plane (Figure \ref{fig:Q_hs_with_cavagnolo}) to that of our sample of FRII radio galaxies. Cygnus A is known to lie in a high density environment, more similar to the cluster environments of the FRI radio galaxies than the group or field environments of typical FRII radio galaxies, and therefore its luminosity is significantly ``environmentally boosted" relative to similar sources located in less dense environments \citep{barthel96}. In Figure \ref{fig:Q_hs_with_cavagnolo} it can be seen that Cygnus A is indeed an outlier compared to our sample of FRII radio galaxies, shifted to higher radio luminosity. This offset is due to the ``environmental boosting" of Cygnus A resulting from its dense environment.  \citet{barthel96} estimate that the luminosity of Cygnus A would be reduced by up to a factor of 30 if it were located in a field environment. This would shift Cygnus A to align with the other FRII radio galaxies in the $Q_{\rm jet} - L_{\rm radio}$ plane. However, Cygnus A is only a factor a few below the FRI $Q_{\rm jet} - L_{\rm radio}$ relation, and so the environmental dependence cannot be the only compensating factor. We note that the effect of ``environmental boosting" in dense environments may be counteracted to some degree by a possible positive correlation between the fraction of energy in non-radiating particles and the environment density \citep{hardcastle10, croston11}. 

Strong shocks driven by powerful FRII radio galaxies may also decrease their radiative efficiency relative to FRI radio galaxies. Unlike FRI radio galaxies, powerful FRII radio galaxies drive strong shocks that sweep up and heat the ambient gas ahead of the cocoon \citep[e.\,g.][]{croston11}. The energy associated with these shocks can be a significant, indeed dominant, factor in the FRII energy budget \citep{worrall12}. 

\citet{willott99} argued that the $Q_{\rm jet} - L_{\rm radio}$ relation should be independent of source size. In Paper~2 we show that this is in fact not the case, due to significant energy losses suffered by radiating electrons though inverse Compton scattering. These losses result in an effective steeping of the spectral index $\alpha$ in Equation~\ref{eqn:k_dependence}, and therefore a decrease in radiative efficiency. For the oldest sources the jet power can easily be underestimated by as much as a factor of three.


Finally, we note that the X-ray cavity jet power estimates used by \citet{cavagnolo10}, and hence the normalisation in Equation \ref{eqn:Q_FRI}, may be underestimated. Shocks, which are currently ignored in the X-ray cavity jet power calculations, may be energetically important in FRI radio galaxies for a substantial part of their evolution, at least in some sources \citep[e.\,g.][]{mcnamara05, fabian06, forman07, wise07, birzan08}. Furthermore, the buoyancy timescale used to estimate the source age could be an overestimate of the true source age, resulting in systematically underestimated jet power measurements \citep{wise07, mcnamara07, birzan08}. 

A somewhat surprising corollary of our results, and the above discussion, is that an extrapolation of the $Q_{\rm jet} - L_{\rm radio}$ relation for low power FRI radio galaxies provides a reasonable approximation for high power sources, despite their vastly different lobe energy budgets.

\subsection{The slope of the $Q_{\rm jet} - L_{\rm radio}$ relation}

The exponent in Equations \ref{eqn:our_best_fit} and \ref{eqn:Q_FRI} is $\lesssim 0.7$. In contrast, the predicted exponent in the $Q_{\rm jet} - L_{\rm radio}$ relation (Equation \ref{eqn:k_dependence}), for a typical spectral index $\alpha \approx 0.8$, is approximately 0.8. 

However, in a flux limited sample, selection bias will cause a systematic increase in radiative efficiency (a decrease in the Willott $f$ factor) with increasing radio luminosity: the low luminosity end will be populated by sources with both low kinetic power and low radiative efficiency (high $f$), while the high luminosity end will be populated by sources with both high kinetic power and high radiative efficiency (low $f$). For this reason alone, it is clear that a single value for the $f$ factor cannot be applied to the entire source population, and the selection bias will flatten (reduce) the observed slope of the $Q_{\rm jet} - L_{\rm radio}$ relation, potentially accounting for the difference between predicted and observed slopes.  

Relatedly, we note that given the common assumption of $f=20$ based on the X-ray cavity jet power measurements for FRI radio galaxies, the Willott et al. relation (Equation \ref{eqn:willott}) predicts that the most luminous FRII radio sources ($L_{151} \gtrsim 3 \times 10^{28}$ W~Hz$^{-1}$~sr$^{-1}$) have jet power in the order of 10$^{48}$ erg~s$^{-1}$. This is equivalent to the Eddington luminosity of a 10$^{10}$ solar mass black hole, and an order of magnitude greater than estimates of jet power in samples of radio galaxies \citep[e.g.][]{rawlings91}. It follows that $f \ll 20$ for the highest luminosity bins. 

\section{Conclusions}

We have presented a new method to measure the kinetic power in AGN jets based on the observed size and luminosity of the jet terminal hotspots. With this new method we were able to confront, from a new perspective, an emerging scenario in which the fraction of energy in non-radiating particles differs between the two morphological classes of radio galaxy (FRI/FRII).  \citet{birzan08} and \citet{croston08} estimate that the ratio of energy in non-radiating particles to the energy in radiating particles, $k$, could be as high as 4000 in some sources, with a median value of approximately 200. In contrast, \citet{croston04, croston05} demonstrate that $k \lesssim 1$ in FRII radio galaxy lobes.  Such a large difference in the lobe energy budgets suggests a large difference between the radiative efficiency of the two morphological classes, of more than an order of magnitude. To test this hypothesis, we estimated the jet kinetic power using the new method based on observed hotspot parameters, for a carefully selected sample of FRII radio galaxies. We compared the resulting $Q_{\rm jet} - L_{\rm radio}$ relation to that determined for FRI radio galaxies by \citet{cavagnolo10} based on jet power measurements determined using the X-ray cavities method. 

We find approximate agreement between the FRI and FRII radio galaxy $Q_{\rm jet} - L_{\rm 151}$ relations, which is ostensibly difficult to reconcile with the differing lobe energy budgets. However, a combination of environmental factors, spectral ageing and strong shocks driven by powerful FRII radio galaxies reduces the radiative efficiency of these objects relative to FRIs, and conspires to move them onto the FRI $Q_{\rm jet} - L_{\rm 151}$ relation. An unexpected outcome of our work is that an extrapolation of the $Q_{\rm jet} - L_{\rm radio}$ relation determined for low power FRI radio galaxies provides a reasonable approximation for high power sources, despite their apparently different lobe energy budgets.

\appendix

\section{The validity of the assumption $F_{\rm E, jet} \approx c \times  F_{\rm M, jet}$ }
 
In this section we consider the requirements for the jet Lorentz factor in order to apply the assumption $F_{\rm E, jet} \approx c \times  F_{\rm M, jet}$.  Consider the ratio
\begin{eqnarray}
\frac{c \times  F_{\rm M, jet}}{F_{\rm E, jet} } &=& \beta + \frac{1}{\Gamma^2 \beta} \left( \frac{p +\frac{B_{\perp}^2 - B_{||}^2}{8 \pi}}{ w + \frac{B_{\perp}^2}{4 \pi}} \right) 
\end{eqnarray}
Let us assume a tangled magnetic field (isotropic distribution in solid angle) in the jet. Then $\langle B_\perp^2 - B_{||}^2  \rangle = (1/3) B^2$ and $\langle B_\perp^2  \rangle = (2/3) B^2$. In that case the above expression can be re-written as follows. 
\begin{eqnarray}
\frac{c \times  F_{\rm M, jet}}{F_{\rm E, jet} } &=& \beta + \frac{1}{\Gamma^2 \beta} \cdot \frac{\frac{1}{3} + \frac{2}{3} \frac{\epsilon_p}{\epsilon_{e^\pm}} + \frac{1}{3} \frac{\epsilon_B}{\epsilon_{e^\pm}}}{ \frac{4}{3} + \frac{5}{3} \frac{\epsilon_p}{\epsilon_{e^\pm}} + \frac{4}{3} \frac{\epsilon_B}{\epsilon_{e^\pm}} + \frac{\rho c^2}{\epsilon_{e^\pm}}} \label{eqn:c_F_M_on_F_E} \\
&=& \beta + \frac{1}{\Gamma^2 \beta} \frac{1}{\xi} 
\end{eqnarray}
In general, provided $\rho c^2 \lesssim \epsilon_{e^\pm}$, we have $\xi \approx 2 - 4$, regardless of the relative energy densities of leptons, protons and magnetic field. However, if $\rho c^2 >> \epsilon_{e^\pm}$, we have $\xi >> 1$, in which case $\frac{c \times  F_{\rm M, jet}}{F_{\rm E, jet} } \approx \beta$. 

Below, we plot the ratio $c\times F_{\rm M, jet}/F_{\rm E, jet}$ as a function of bulk Lorentz factor $\Gamma$ using Equation \ref{eqn:c_F_M_on_F_E}, for various combinations of the parameters $\frac{\epsilon_p}{\epsilon_{e^\pm}}, \frac{\epsilon_B}{\epsilon_{e^\pm}}, \frac{\rho c^2}{\epsilon_{e^\pm}}$.

\begin{figure}[!h]
\epsscale{0.85}
\plotone{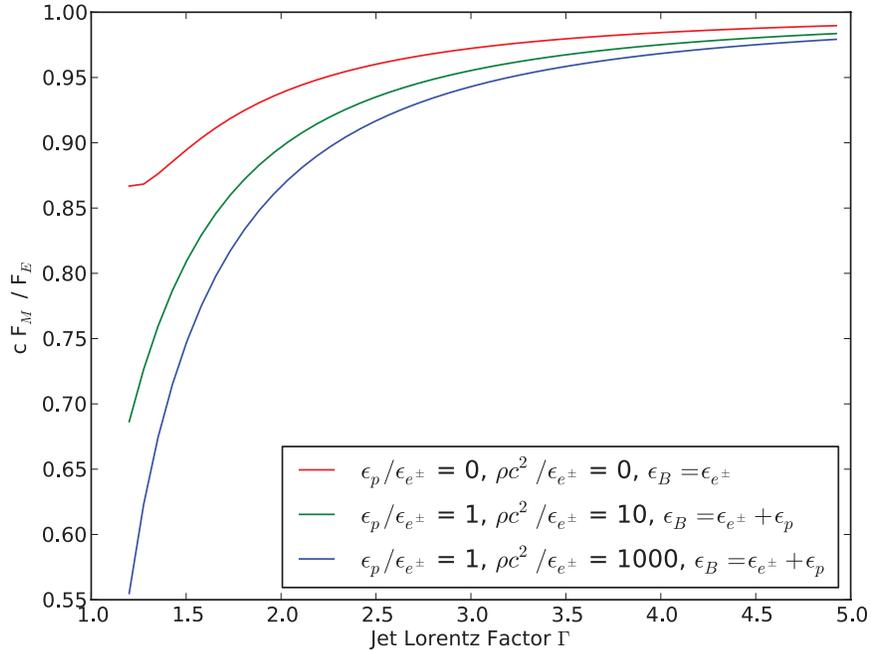}
\caption{Plot of the ratio of speed of light times the momentum flux in the jet to the energy flux as a function of jet Lorentz factor for a range of jet plasma conditions. \label{fig:c_F_M_on_F_E}}
\end{figure}

\citet{mullin09} found that the radio emitting plasma in kpc-scale radio jets has a characteristic bulk Lorentz factor in the range $\Gamma = 1.18 - 1.49$. However, inverse Compton modelling of kpc-scale quasar X-ray jets suggests bulk Lorentz factors in the order of $\Gamma \sim 10$ \citep[see e.\,g.][]{kataoka05}. \citet{mullin09} argue that this discrepancy between jet speed estimates indicates the need for velocity structure across the jet (i.e. spine-sheath type of model). In that case, the average Lorentz factor of jet material will be $\Gamma > 1.2 - 1.5$, and potentially much greater. Figure \ref{fig:c_F_M_on_F_E} illustrates that, at worst, the value of $c \times F_M$ will underestimate the true jet power by up to a few tens of percent. However, the most likely scenario is that  $c \times F_M$ will be within a few percent of the jet power. Note that the estimate will be an underestimate, so that this method represents a lower limit to the true jet power.

\acknowledgments 

L.E.H.G. is grateful to Geoff Bicknell for enlightening discussions on the use of radio galaxy hotspots for jet power measurements. We thank the anonymous referee for useful comments that helped us to improve the paper. 



\begin{thebibliography}{}

\bibitem[Barthel 
\& Arnaud(1996)]{barthel96} Barthel, P.~D., \& Arnaud, K.~A.\ 1996, \mnras, 283, L45 


\bibitem[Belsole et al.(2007)]{belsole07} Belsole, E., Worrall, D.~M., Hardcastle, M.~J., \& Croston, J.~H.\ 2007, \mnras, 381, 1109 


\bibitem[B{\^i}rzan et al.(2008)]{birzan08} B{\^i}rzan, L., McNamara, B.~R., Nulsen, P.~E.~J., Carilli, C.~L., 
\& Wise, M.~W.\ 2008, \apj, 686, 859 

\bibitem[Blundell et al.(1999)]{blundell99} Blundell, K.~M., 
Rawlings, S., \& Willott, C.~J.\ 1999, \aj, 117, 677 


\bibitem[Carilli et al.(1991)]{carilli91} Carilli, C.~L., Perley, 
R.~A., Dreher, J.~W., \& Leahy, J.~P.\ 1991, \apj, 383, 554 \bibitem[Cattaneo \& Best(2009)]{cattaneo09} Cattaneo, A., \& Best, P.~N.\ 2009, \mnras, 395, 518 


\bibitem[Cavagnolo et al.(2010)]{cavagnolo10} Cavagnolo, K.~W., 
McNamara, B.~R., Nulsen, P.~E.~J., et al.\ 2010, \apj, 720, 1066 


\bibitem[Croston et al.(2003)]{croston03} Croston, J.~H., Hardcastle, M.~J., Birkinshaw, M., \& Worrall, D.~M.\ 2003, \mnras, 346, 1041 

\bibitem[Croston et al.(2004)]{croston04} Croston, J.~H., Birkinshaw, M., Hardcastle, M.~J., \& Worrall, D.~M.\ 2004, \mnras, 353, 879 

\bibitem[Croston et al.(2005)]{croston05} Croston, J.~H., Hardcastle, M.~J., Harris, D.~E., et al.\ 2005, \apj, 626, 733 

\bibitem[Croston et al.(2008)]{croston08} Croston, J.~H., Hardcastle, M.~J., Birkinshaw, M., Worrall, D.~M., \& Laing, R.~A.\ 2008, \mnras, 386, 1709 


\bibitem[Croston et al.(2011)]{croston11} Croston, J.~H., Hardcastle, M.~J., Mingo, B., et al.\ 2011, \apjl, 734, L28 



\bibitem[Croton et al.(2006)]{croton06}
Croton, D.\ J., Springel, V., White, S.\ D.\ M., et al. 2006, MNRAS, 365, 11




\bibitem[Daly(2009)]{daly09} Daly, R.~A.\ 2009, \apjl, 691, 
L72 



\bibitem[Daly et al.(2012)]{daly12} Daly, R.~A., Sprinkle, 
T.~B., O'Dea, C.~P., Kharb, P., \& Baum, S.~A.\ 2012, \mnras, 423, 2498 



\bibitem[Dennett-Thorpe et al.(1997)]{dennett-thorpe97} Dennett-Thorpe, 
J., Bridle, A.~H., Scheuer, P.~A.~G., Laing, R.~A., 
\& Leahy, J.~P.\ 1997, \mnras, 289, 753 


\bibitem[De Young(2002)]{deyoung02} De Young, D.~S.\ 2002, New Astronomy Reviews, 
46, 393

\bibitem[Double et al.(2004)]{double04} Double, G.~P., Baring, 
M.~G., Jones, F.~C., \& Ellison, D.~C.\ 2004, \apj, 600, 485

\bibitem[Fabian et al.(2006)]{fabian06} Fabian, A.~C., Sanders, J.~S., Taylor, G.~B., et al.\ 2006, \mnras, 366, 417 

\bibitem[Falle(1991)]{falle91} Falle, S.~A.~E.~G.\ 1991, 
\mnras, 250, 581


\bibitem[Fanidakis et al.(2011)]{fanidakis11}
{Fanidakis}, N., {Baugh}, C.~M., {Benson}, A.~J., {Bower}, R.~G., {Cole}, S., {Done}, C., {Frenk}, C.~S. 2012, MNRAS, 410, 53


\bibitem[Fernandes et al.(2011)]{fernandes11} Fernandes, C.~A.~C., Jarvis, M.~J., Rawlings, S., et al.\ 2011, \mnras, 411, 1909 


\bibitem[Forman et al.(2007)]{forman07} Forman, W., Jones, C., Churazov, E., et al.\ 2007, \apj, 665, 1057 



\bibitem[Georganopoulos 
\& Kazanas(2003)]{georganopoulos03} Georganopoulos, M., \& Kazanas, D.\ 2003, \apjl, 589, L5 

\bibitem[Godfrey et al.(2009)]{godfrey09} Godfrey, L.~E.~H., et 
al.\ 2009, \apj, 695, 707 

\bibitem[Godfrey et al.(2012)]{godfrey12} Godfrey, L.~E.~H., Bicknell, G.~V., Lovell, J.~E.~J., et al.\ 2012, \apj, 755, 174 

\bibitem[Hardcastle et al.(1998)]{hardcastle98} Hardcastle, M.~J., Alexander, P., Pooley, G.~G., \& Riley, J.~M.\ 1998, \mnras, 296, 445 

\bibitem[Hardcastle et al.(2001a)]{hardcastle01a} Hardcastle, M.~J., 
Birkinshaw, M., \& Worrall, D.~M.\ 2001, \mnras, 323, L17 

\bibitem[Hardcastle(2001b)]{hardcastle01b} Hardcastle, M.~J.\ 2001b, \aap, 373, 881 

\bibitem[Hardcastle(2003)]{hardcastle03} Hardcastle, M.~J.\ 2003, New Astronomy Reviews, 47, 649 

\bibitem[Hardcastle et al.(2004)]{hardcastle04} Hardcastle, M.~J., 
Harris, D.~E., Worrall, D.~M., \& Birkinshaw, M.\ 2004, ApJ, 612, 729 

\bibitem[Hardcastle et al.(2007)]{hardcastle07} Hardcastle, M.~J., 
Evans, D.~A., \& Croston, J.~H.\ 2007, \mnras, 376, 1849 

\bibitem[Hardcastle et al.(2009)]{hardcastle09} Hardcastle, M.~J., Evans, D.~A., \& Croston, J.~H.\ 2009, \mnras, 396, 1929 


\bibitem[Hardcastle \& Croston(2010)]{hardcastle10} Hardcastle, M.~J., \& Croston, J.~H.\ 2010, \mnras, 404, 2018 



\bibitem[Harris et al.(1994)]{harris94} Harris, D.~E., Carilli, 
C.~L., \& Perley, R.~A.\ 1994, \nat, 367, 713 

\bibitem[Harris et al.(2000)]{harris00} Harris, D.~E., et al.\ 
2000, \apjl, 530, L81

\bibitem[Ito et al.(2008)]{ito08} Ito, H., Kino, M., 
Kawakatu, N., Isobe, N., \& Yamada, S.\ 2008, \apj, 685, 828 

\bibitem[Kaiser 
\& Alexander(1997)]{kaiser97} Kaiser, C.~R., \& Alexander, P.\ 1997, \mnras, 286, 215 

\bibitem[Kataoka 
\& Stawarz(2005)]{kataoka05} Kataoka, J., \& Stawarz, {\L}.\ 2005, \apj, 622, 797 


\bibitem[Lazio et al.(2006)]{lazio06} Lazio, T.~J.~W., Cohen, 
A.~S., Kassim, N.~E., Perley, R.~A., Erickson, W.~C., Carilli, C.~L., \& 
Crane, P.~C.\ 2006, \apjl, 642, L33 


\bibitem[Laing(1989)]{laing89} Laing, R.\ 1989, Hot Spots in Extragalactic Radio Sources, 327, 27 


\bibitem[Lobanov(1998)]{lobanov98} Lobanov, A.~P.\ 1998, \aap, 330, 79 


\bibitem[Mart{\'{\i}}nez-Sansigre \& Rawlings(2011)]{martinez-sansigre11} Mart{\'{\i}}nez-Sansigre, A., \& Rawlings, S.\ 2011, \mnras, 414, 1937 



\bibitem[McNamara et al.(2005)]{mcnamara05} McNamara, B.~R., Nulsen, P.~E.~J., Wise, M.~W., et al.\ 2005, \nat, 433, 45 



\bibitem[McNamara \& Nulsen(2007)]{mcnamara07} McNamara, B.~R., \& Nulsen, P.~E.~J.\ 2007, \araa, 45, 117 


\bibitem[Miller et al.(1999)]{miller99} Miller, N.~A., Owen, 
F.~N., Burns, J.~O., Ledlow, M.~J., \& Voges, W.\ 1999, \aj, 118, 1988 



\bibitem[Mullin et al.(2008)]{mullin08} Mullin, L.~M., Riley, 
J.~M., \& Hardcastle, M.~J.\ 2008, \mnras, 390, 595 


\bibitem[Mullin 
\& Hardcastle(2009)]{mullin09} Mullin, L.~M., \& Hardcastle, M.~J.\ 2009, \mnras, 398, 1989 



\bibitem[O'Sullivan et al.(2011)]{osullivan11} O'Sullivan, E., 
Giacintucci, S., David, L.~P., et al.\ 2011, \apj, 735, 11 


\bibitem[Rafferty et al.(2006)]{rafferty06} Rafferty, D.~A., 
McNamara, B.~R., Nulsen, P.~E.~J., \& Wise, M.~W.\ 2006, \apj, 652, 216 


\bibitem[Rawlings \& Saunders(1991)]{rawlings91} Rawlings, S., \& Saunders, R.\ 1991, \nat, 349, 138 


\bibitem[Rawlings \& Jarvis(2004)]{rawlings04} Rawlings, S., Jarvis, M. J.  2004, MNRAS Lett., 355, 9


\bibitem[Saxton et al.(2002)]{saxton02} Saxton, C.~J., Sutherland, R.~S., Bicknell, G.~V., Blanchet, G.~F., \& Wagner, S.~J.\ 2002, \aap, 393, 765 


\bibitem[Saxton et al.(2010)]{saxton10} Saxton, C.~J., Wu, K., 
Korunoska, S., et al.\ 2010, \mnras, 405, 1816 

\bibitem[Shabala \& Alexander(2009)]{shabala09} Shabala, S.~S., Alexander, P. 2009, ApJ, 699, 525

\bibitem[Shabala et al.(2011)]{shabala11} Shabala, S. S., Kaviraj, S., Silk, J. 2011, MNRAS, 413, 2815

\bibitem[Shabala et al.(2012)]{shabala12} Shabala, S.~S., Santoso, J.~S., \& Godfrey, L.~E.~H.\ 2012, \apj, 756, 161 

\bibitem[Paper 2(2012)]{paper2} Shabala, S.~S., \& Godfrey, L.~E.~H.\ 2012, submitted to \apj

\bibitem[Simpson(1998)]{simpson98} Simpson, C.\ 1998, \mnras, 297, L39 


\bibitem[Tadhunter et al.(1998)]{tadhunter98} Tadhunter, C.~N., Morganti, R., Robinson, A., et al.\ 1998, \mnras, 298, 1035 


\bibitem[Willott et al.(1999)]{willott99} Willott, C.~J., 
Rawlings, S., Blundell, K.~M., \& Lacy, M.\ 1999, \mnras, 309, 1017 


\bibitem[Wilson et al.(2006)]{wilson06} Wilson, A.~S., Smith, 
D.~A., \& Young, A.~J.\ 2006, \apjl, 644, L9 


\bibitem[Wise et al.(2007)]{wise07} Wise, M.~W., McNamara, B.~R., Nulsen, P.~E.~J., Houck, J.~C., \& David, L.~P.\ 2007, \apj, 659, 1153 




\bibitem[Worrall \& Birkinshaw(2006)]{worrall06} Worrall, D.~M., \& Birkinshaw, M.\ 2006, Physics of Active Galactic Nuclei at all Scales, 693, 39 



\bibitem[Worrall(2009)]{worrall09} Worrall, D.~M.\ 2009, A\&ARv, 17, 1 


\bibitem[Worrall et al.(2012)]{worrall12} Worrall, D.~M., Birkinshaw, M., Young, A.~J., et al.\ 2012, \mnras, 424, 1346 



\bibitem[Wright 
\& Birkinshaw(2004)]{wright04} Wright, M.~C.~H., \& Birkinshaw, M.\ 2004, \apj, 614, 115 


\bibitem[Zirbel(1997)]{zirbel97} Zirbel, E.~L.\ 1997, \apj, 476, 
489 



\end{thebibliography}
\end{document}